\documentclass[aps,prl,twocolumn,showpacs,amsmath,amssymb,eqnarray,array]{revtex4}
\usepackage[usenames]{color}
\usepackage{graphicx}
\usepackage{dcolumn}
\usepackage{bm}
\begin{document}

\title{Exploring the magnetoroton excitations in quantum wires: Inverse dielectric functions\\}
\author{Manvir S. Kushwaha$^*$}
\address
{\centerline{Institute of Industrial Science, The University of Tokyo, 4-6-1 Komaba, Meguro-Ku,
Tokyo 153-8505, Japan}}

\begin{abstract}

A theoretical investigation has been made of the magnetoplasmon excitations in a quantum wire characterized
by a confining harmonic potential and subjected to a perpendicular magnetic field. We study the (nonlocal,
dynamic) inverse dielectric function to examine the charge-density excitations within a two-subband model
in the framework of Bohm-Pines' random-phase approximation. A particular stress is put on the (intersubband)
magnetoroton excitation which changes the sign of its group velocity twice before merging with the respective single-particle continuum. It has already been suggested that the electronic device based on such magnetoroton excitations can act as an {\it active} laser medium [see, e.g., Phys. Rev. B {\bf 78}, 153306 (2008)].
Scrutinizing the real and imaginary parts of the inverse dielectric function provides us with an important
information on the longitudinal and transverse (Hall) resistances of the system.

\end{abstract}


\date{\today}

\maketitle


A quantum wire [or, more realistically, quasi-one dimensional electron gas (Q1DEG) for better and broader range
of physical understanding] lies in the middle of the rainbow composed of quasi-two, quasi-one, and quasi-zero
dimensional electron systems. The quantum wires are by now well-known to possess some very unique transport
properties such as magnetic depopulation, electron waveguide, quenching of the Hall effect, quantization of
conductance, negative energy dispersion, magnetoroton excitations, and spin-charge separation, to name a few.
These are well documented in the literature (see, for instance, Ref. 1). It is also well-established that the
realistic quantum wires are safely describable as Fermi liquids rather than Luttinger liquids$^2$ and hence
the Fermi-liquid-like Bohm-Pines' random-phase approximation (RPA) is justifiable by all means.

Here we aim at investigating the charge-density excitations in a realistic quantum wire within a two-subband
model in the framework of Bohm-Pines' RPA. The main focus of this study will be the (intersubband) magnetoroton excitation which changes the sign of its group velocity twice before merging with the respective single-particle continuum. A roton is an elementary excitation whose dispersion relation shows a linear increase from the origin,
but exhibits first a maximum, and then a minimum in energy as the momentum increases. Excitations with momenta
in the linear region are called phonons; those with momenta near the maximum are called maxons; and those with
momenta close to the minimum are called rotons.

A magnetoroton mode in Q1DEG was predicted within the framework of Hartree-Fock approximation in 1992 [see Ref. 3]
and it was soon verified in the resonant Raman scattering experiments in 1993 [see Ref. 4]. However, a theoretical finding of the magnetoroton mode in realistic quantum wires had been elusive$^{5,6}$ until very recently. One of
the reasons behind such elusion has been the overplay with the theory, as was done in Refs. 5 and 6, which both
missed to observe this magnetoroton mode.

Recently we embarked on the detailed investigation of the magnetoroton (MR) excitations and their characteristic dependence on the magnetic field, the charge density, and the confinement potential$^7$. There it was illustrated
that the MR exhibits negative group velocity between the maxon maximum and roton minimum. \textcolor{blue}
{Consequently, it leads to the tachyon-like (superluminal) behavior$^8$ without one's having to introduce negative energies}. The interest in negative group velocity is based on anomalous dispersion in a gain medium, where the sign
of the phase velocity is the same for incident and transmitted waves and energy flows inside the gain medium in the opposite direction to the incident energy flow in vacuum. The insight is that demonstration of negative group
velocity is possible in media with inverted populations, so that gain instead of absorption occurs at the frequencies
of interest. \textcolor{blue}{It is tempting to know that such a state with inverted population can be characterized
by {\em negative temperature}}. A medium with an inverted population has the remarkable ability of amplifying a small optical signal of definite wavelength, i.e., it can serve as an {\em active} laser medium.

In the present note, we analyze the inverse dielectric function (IDF) of a Q1DEG and correlate the real and imaginary parts thereof to the longitudinal and transverse (Hall) resistances of the system. For this purpose, we consider a
Q1DEG in the $x-y$ plane with a parabolic confining potential $V(x)=\frac{1}{2}m^*\omega_0^2 x^2$ along the $x$ direction and a magnetic field applied along the $z$ direction in the Landau gauge [$\bf {A}=(0,Bx,0)$]. Here
$\omega_0$ is the characteristic frequency of the harmonic oscillator and $m^*$ the electron effective mass of the system. The resultant system is a realistic quantum wire with free propagation along the $y$ direction and the {\em magnetoelectric} quantization along $x$ direction. In general, we aim at computing the inverse dielectric function for the magnetized quantum wires in the framework of Bohm-Pines' RPA within a two-subband model, with only the lowest one assumed to be occupied.

The study of the inverse dielectric function leads us to substantiate the magnetoplasmon excitations usually searched
with the zeros of the dielectric function. This is what we should anticipate because the zeros of the dielectric
function and the poles of the inverse dielectric function must yield exactly identical results. A careful analysis of the inverse (nonlocal, dynamic) dielectric function [$\epsilon^{-1}(q_y,\omega)$] closely reflects on the longitudinal and transverse (Hall) resistances in such realistic quantum wires. For instance, the longitudinal (transverse) resistance $\rho_{yy}$ ($\rho_{xy}$) is determined by the imaginary (real) part of the inverse dielectric function. Moreover, the imaginary part of the inverse dielectric function also furnishes a very significant estimates of the
Raman (and electron) scattering cross-section. For the extensive formalism on the inverse dielectric functions (for quasi-n dimensional systems), a reader is referred to Ref. 1.

As to the illustrative numerical examples, we focus on the {\em narrow} channels of the In$_{1-x}$Ga$_x$As system (see, e.g., Ref. 7). We compute the magnetoplasmon excitations in a Q1DEG in the framework of RPA within a two-subband model in the presence of a perpendicular magnetic field $B$ at T=0 K. We do so by examining the influence of several parameters involved in the analytical results. These are, for instance, the 1D charge density $n_{1d}$,
characteristic frequency of the harmonic potential $\omega_o$, and the magnetic field $B$. The material parameters
used are: effective mass $m^*=0.042 m_{_0}$, the background dielectric constant $\epsilon_{_b}=13.9$, 1D charge
density $n_{1D}=1.0\times 10^{6}$ cm$^{-1}$, confinement energy $\hbar\omega_0=2.0$ meV, and the effective
confinement width of the harmonic potential well, estimated from the extent of the Hermite function, $w_{eff}=40.19$ nm. Notice that the Fermi energy $\epsilon_F$ varies in the case where $n_{1D}$, $B$, or $\hbar \omega_{0}$ is varied.

\begin{figure}[htbp]
\includegraphics*[width=8cm,height=9cm]{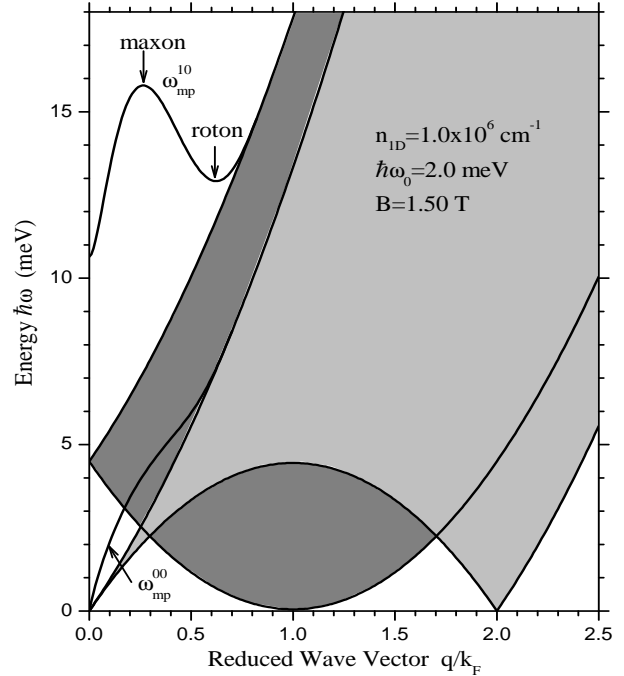}
\caption{(Color Online) The full magnetoplasmon dispersion plotted as energy $\hbar \omega$ vs. reduced wave vector
$q/k_F$. The light (dark) shaded region refers to the intrasubband (intersubband) single-particle  excitations. The
bold curve marked as $\omega^{00}_{mp}$ ($\omega^{10}_{mp}$) stands for the intrasubband (intersubband) collective (magnetoplasmon) excitations. The other parameters are as given inside the picture.}
\label{fig1}
\end{figure}

Figure 1 illustrates the full magnetoplasmon dispersion for the  Q1DEG plotted as energy $\hbar \omega$ vs
reduced wave vector $q/k_F$, for the given values of $n_{1d}$, $\hbar\omega_0$, and $B$. The light (dark)
shaded region refers, respectively, to the intrasubband (intersubband) single-particle excitations (SPE).
The bold solid curve marked as $\omega^{00}_{mp}$ ($\omega^{10}_{mp}$) is the intrasubband (intersubband)
collective (magnetoplasmon) excitations (CME). The intrasubband CME start from the origin and merges
with the upper edge of the intrasubband SPE at ($q_y/k_F=0.66$, $\hbar\omega=7.68$ meV)  and thereafter
ceases to exist as a bonafide, long-lived CME. The intersubband CME starts at ($q_y/k_f=0$,
$\hbar\omega=10.65$ meV), attains a maximum at ($q_y/k_f=0.28$, $\hbar\omega=15.80$ meV), reaches its
minimum at ($q_y/k_f=0.63$, $\hbar\omega=12.93$ meV), and then rises up to merge with the upper edge of
the intersubband SPE at ($q_y/k_f=0.82$, $\hbar\omega=14.63$ meV). It is a simple matter to check
(analytically) why the energy of the lower branch of the intrasubband SPE goes to zero at $q_y=2k_F$, why
the lower branch of the intersubband SPE observes its {\em minimum} ({\em not zero}) at $q_y=k_F$, and why
the intersubband SPE starts at the subband spacing ($\hbar\tilde{\omega}=4.4915$ meV) at the origin. The
most interesting aspect of this excitation spectrum is the existence of this intersubband CME (referred to
as the magnetoroton). \textcolor{blue}{Notice how the magnetoroton mode changes the sign of its group velocity
twice before merging with the respective SPE}. At $q_y=0$, the energy difference between the intersubband CME
and SPE is a manifestation of the many-body effects such as depolarization and excitonic shifts$^{1}$.

\begin{figure}[htbp]
\includegraphics*[width=8cm,height=9cm]{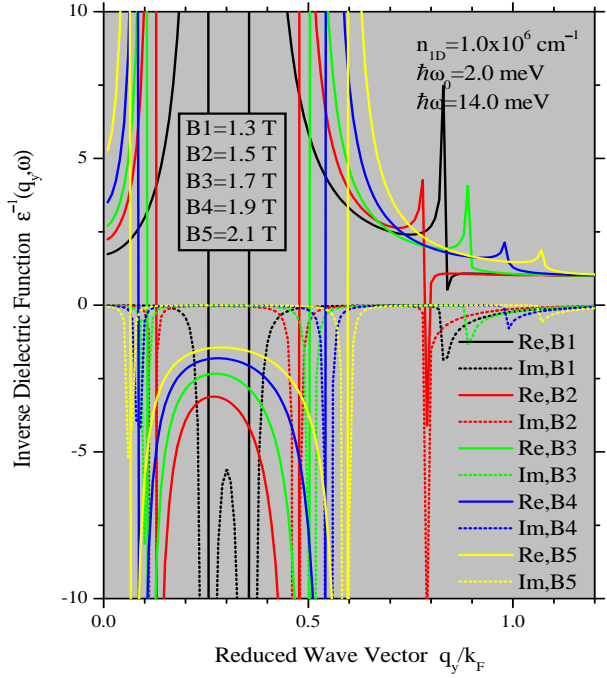}
\caption{(Color Online) Inverse dielectric function $\epsilon^{-1}(q_y, \omega)$ vs. reduced wave vector
$q_y/k_F$ for various values of the magnetic field ($B$). The other parameters are as given inside the picture.}
\label{fig2}
\end{figure}

Figure 2 illustrates the inverse dielectric function $\epsilon^{-1}(q_y, \omega)$ as a function of the reduced propagation vector $q_y/k_F$ for various values of the magnetic field ($B$). Notice that the excitation energy
is fixed at $\hbar\omega=14.0$ meV. It is noteworthy that the quantity that directly affects the transport
phenomenon is the spectral weight Im[$\epsilon^{-1}(q_y, \omega)$], which contains both the single-particle
contribution at large propagation vector ($q_y$) and the collective (magnetoplasmon) contribution at small $q_y$.
It is clearly observed that, for $B=0.21$ T, the sharp peaks at the reduced wave vector $q_y/k_F\simeq 0.055$ and
0.596 stand for the two positions of the magnetoroton for the given energy $\hbar\omega=14.0$ meV: the first
corresponds to the phononic part and the second to the rotonic one. Similarly, the broader
peak at $q_y/k_F\simeq 1.069$ refers to the respective intersubband single-particle excitations (SPE). The latter actually stands for the position where the roton mode merges with the respective intersubband SPE. The other
peaks for different values of the magnetic field $B$ can be clearly identified. \textcolor{blue}{Clearly, these
peaks are a result of the existing poles of the inverse dielectric function in the ($\omega-q_y$) space}.

\begin{figure}[htbp]
\includegraphics*[width=8cm,height=9cm]{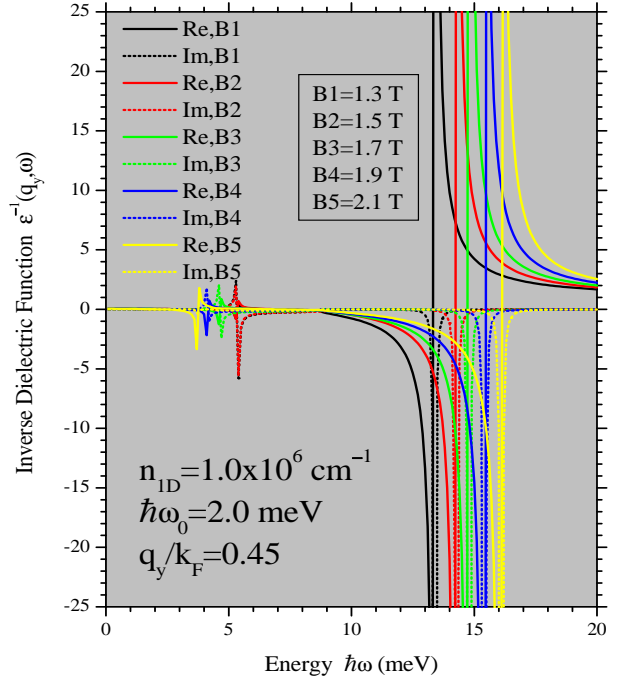}
\caption{(Color Online) Inverse dielectric function $\epsilon^{-1}(q_y, \omega)$ vs. the energy $\hbar \omega$ for various values of the magnetic field ($B$). The other parameters are as given inside the picture.}
\label{fig3}
\end{figure}

Figure 3 illustrates the inverse dielectric function $\epsilon^{-1}(q_y, \omega)$ as a function of the excitation  energy $\hbar\omega$ for various values of the magnetic field ($B$). The reduced propagation vector is fixed
at $q_y/k_F=0.45$. Again, in analogy with the previous discussion, we focus on the imaginary part of the inverse
dielectric function. Since we choose a relatively small values of the propagation vector, the peaks in this figure
can only identify the collective (magnetoplasmon) excitations. For the magnetic field $B=2.1$ T, we notice that
the lower peak at $\hbar\omega \simeq 3.737$ meV stands for the intrasubband magnetoplasmon in the quantum wire,
whereas the higher peak at $\hbar\omega \simeq 16.059$ meV corresponds to the rotonic part of the intersubband magnetoplasmon, which is identified as the magnetoroton mode in the corresponding system. Other peaks for
different values of the magnetic field can be similarly identified. For such a smaller value of the propagation
vector, no resonance peak corresponds to any single-particle excitation spectrum in the system.

It is interesting to notice the occurrence of the resonance peaks in both figures with the variation of the
magnetic field. Figure 2 shows that the lower (higher) peak of the magnetoroton mode occurs at lower
(higher) values of the propagation vector as the magnetic field is enhanced. \textcolor{blue}{Also, generally
speaking, the higher the magnetic field, the larger the propagation vector for the resonance peaks corresponding
to the single-particle excitations. Figure 3 depicts that the higher the magnetic field, the smaller (larger) the
excitation energy of the intrasubband (intersubband, or magnetoroton) mode. This leads us to infer
that the higher the magnetic field, the greater the energy of the magnetoroton mode in the ($\omega-q_y$)
space of the spectrum}. A similar conclusion can be drawn by having a closer look at Fig. 1 in Ref. 7, where we
had, in fact, searched the zeros of the dielectric function. Another interesting aspect is that the resonance
peaks associated with the magnetoroton mode are sharper and greater (in magnitude) than those related with the
intrasubband magnetoplasmons and/or the single-particle excitations. This remark is equally valid for both Figs.
2 and 3.

In summary, we have investigated the inverse dielectric functions for a realistic quantum wire in the presence of
a confining harmonic potential and subjected to a quantizing perpendicular magnetic field. We focused on the charge-density excitations within a two-subband model in the framework of Bohm-Pines' full RPA. While the poles of
the inverse dielectric function yield exactly the same collective (magnetoplasmon) and single-particle excitations
as obtained by searching the zeros of the dielectric function, the former certainly has remarkable advantages over
the latter. \textcolor{blue}{For instance, the imaginary (real) part of the inverse dielectric function turns out
to yield a significant measure of the longitudinal (Hall) resistance in the system. This implies that exploring the
IDF paves the way to understand not only the optical but also the transport phenomena in the system of interest}. Furthermore, the quantity Im[$\epsilon^{-1}(q_y, \omega)$] also implicitly provides the details of the inelastic
light (or electron) scattering cross-section $S(q_y)$ in the system. That means that the longitudinal resistance $\rho_{yy}$ can also be understood as a weighted sum of the scattering cross-section. As to the inelastic electron scattering, Im[$\epsilon^{-1}(q_y, \omega)$] plays a direct role in computing, for example, such important phenomenon
as the fast-particle energy loss to a thin film, quantum well, or quantum wire, ...etc.

Finally, we recall that the existence of the magnetoroton mode in quantum wires is solely attributed to the applied magnetic field. However, it is observed that there is a minimum (threshold) value of the magnetic field below which
the MR does not exist. The roton minimum is the mode of highest density of excitation states. The electron device
based on the MR (associated with the negative group velocity) can act as an {\em active} laser medium. \textcolor{blue}{The notion of negative group velocity, anomalous dispersion, gain (rather than absorption) band, inverted population, negative temperature, and superluminal behavior is so very intricate and important to practical
realization of the {\em active} laser medium}. The MR features are among the most significant manifestations of the many-body effects. Currently, we are investigating such many-particle interactions with respect to the MR and the results will be reported shortly.

\begin{acknowledgments}
During the course of this work I have benefited from many useful discussions and communications with some colleagues.
I would particularly like to thank Hiroyuki Sakaki, Godfrey Gumbs, and Bahram Djafari-Rouhani.
\end{acknowledgments}




\end{document}